\documentclass{ws-procs9x6}            
\begin{document}
\title{Solar neutrino flare, megaton neutrino detectors and human space journey}

\author{Daniele Fargion{$^*$}{$^{a,}$}{$^c$}, Pietro Oliva {$^{b,}$}{$^{c,}$}{$^d$},{{P.~G.} {De Sanctis Lucentini}}$^{~f}$, Silvia Pietroni {$^a$}, Fabio La Monaca {$^a$}, Paolo Paggi{$^a$} ,Emanuele Habib{$^h$},  Maxim Yu. Khlopov {$^e$}}

\address{
{$^a$} Physics Department Rome University 1,\\
P.le A. Moro 2, 00185, Rome, Italy\\
{$^b$} Science Department Roma Tre University,\\ Via della Vasca Navale, 84, 00146, Rome, Italy\\
{$^c$} MIFP, Via Appia Nuova 31, 00040 Marino (Rome), Italy\\
$^d$   Niccol\`o Cusano University,\\ Via Don Carlo Gnocchi 3, 00166 Rome, Italy\\
{$^f$}  Physics Department, Gubkin Russian State University (National Research
      University), 65 Leninsky Prospekt, Moscow, 119991, Russian Federation\\
{$^h$} DIAEE, Rome University, Via Eugossiana 18,00184, Rome Italy \\
{$^e$}  Institute of Physics, Southern Federal University\\
Stachki 194, Rostov on Don 344090, Russia
{$^*$}E-mail: daniele.fargion@roma1.infn.it
}

\begin{abstract}

The largest solar flares have been recorded in gamma flash and hard spectra up to tens GeV energy. The present building and upgrade of Hyper-Kamiokande (HK) in Japan and Korea, (as well as Deep Core, PINGU) Megatons neutrino detectors do offer a novel way to detectable traces of solar flares: their sudden anti-neutrino (or neutrino) imprint made by proton scattering and pion decays via Delta resonance production on solar corona foot-point. These signals might be observable at largest flare by HK via soft spectra up to tens-hundred MeV energy and by IceCube-PINGU at higher, GeVs energies. We show the expected rate of signals for the most powerful solar flare occurred in recent decades extrapolated for future Megaton detectors.
The neutrino solar flare detection with its prompt alarm system may alert astronauts on space journey allowing them to protect themselves into inner rocket containers surrounded by fuel or water supply. These container walls are able to defend astronauts  from the main lethal (the dominant soft component) radiation wind due to such largest solar flares.
\end{abstract}


\bodymatter

\section{Introduction}
\begin{figure}[!t]
\begin{center}
\includegraphics[width=0.9\textwidth]{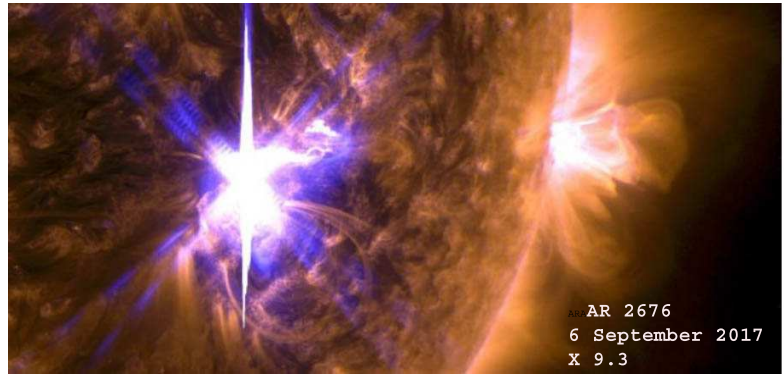}
\caption{The last year solar flare on 6\textsuperscript{th} September at 11.53-12.02 UT. It was able to shine huge amount of x-rays (X 9.3) and a delayed proton wind capable to disturb or even disrupt several GPS satellite \cite{doi:10.1029/2018SW001897} and to lead also to huge geomagnetic storms and boreal lights \cite{doi:10.1029/2018SW001932}.}\label{Fig0}
\end{center}
\end{figure}

\begin{figure}[!t]
\begin{center}
\includegraphics[width=0.9\textwidth]{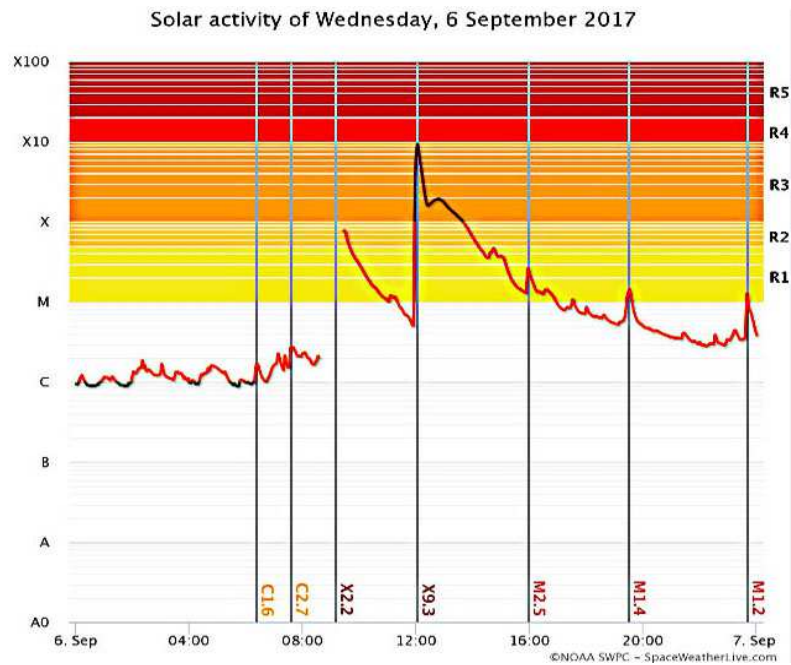}
\caption{The 6\textsuperscript{th} September 2017 flare was able to shine huge amount of x-rays \cite{SF}; it has been the largest solar flare since 2005. Such huge events may take place once every three or five years on average.}\label{Fig0b}
\end{center}
\end{figure}

 Our life survival as well as the main star evolution understanding is based on the Sun radiation  by an internal, steady, thermonuclear conversion mass in energy, fusing hydrogen into helium nuclei: this huge mass conversion into pure radiation shines on us since billions of years (several million of tons each second, mostly $\gamma$ photons and a tiny -- $2.3\%$ -- $\nu$ solar component).
  There are several forms of stored energy (as the gravitational one) on the Sun. A small fraction of the solar energy it is also hidden in the corona by large magnetic field frozen in huge plasma loops; these loops and strings are often dancing, connecting and reconnecting, leading  to rare  huge energy explosions in bright x-ray flare and exceptional mass ejection of solar particle burst, wind or flares. One of the brightest of this event had occurred very recently, just a year ago  on 6\textsuperscript{th} Sepetember 2017 \cite{2018arXiv180307654O}, as here shown in the x-ray figures: see Fig.~\ref{Fig0} and by its time evolution peaks in Fig.~\ref{Fig0b}.

The amount of energy released by largest solar flare may reach several billions of Hiroshima bomb energy.
This huge burst at Sun distance would not be detectable by their (fission nuclear) neutrinos, being diluted mostly at low (few MeV) energy.
This energy power is indeed a very small, below 1\% of the steady inner thermonuclear $p-p$ solar power (at half  MeV energy range); however, it is a huge focalized power burst on the solar corona skin where particle scattering among themselves may produce pion secondaries: their neutrino traces may reach several tens or hundreds MeV energies, possibly reaching Earth in a detectable threshold edge.

The X-ray peak component in brightest solar flare may reach  a part as  1\% of the solar optical fluency, while the kinetic energy of the corona mass emission may be as large as a part over a thousand of the sun luminosity. Nevertheless, the solar largest flare energy spectra is harder than the inner thermonuclear solar ones and it may overcome the cosmic ray flux by three or four order of magnitude, for a few minute time peak. The solar flare particle arriving on Earth may hit our atmosphere and increase, within a prompt or a diluted delayed arrival, the cosmic ray rain. These late, because bent in spiral by magnetic forces, ``Terrestrial''  cosmic winds or solar flare may increase the atmosphere secondaries leading to Delta resonance, whose decay in pions and later in  muons and final neutrinos, are a guaranteed, detectable, but a somehow useless, signal.
 There is also an opposite late effect of the solar flare wind sweeping in a spiral way the common Earth cosmic rays arrival: the so called Forbush  decrease \cite{PhysRev.51.1108.3} that leads not to an increase, but to a detectable delayed decrease or suppression of the cosmic ray fluency, days after the burst.

\begin{figure}[!t]
\begin{center}
\includegraphics[width=0.9\textwidth]{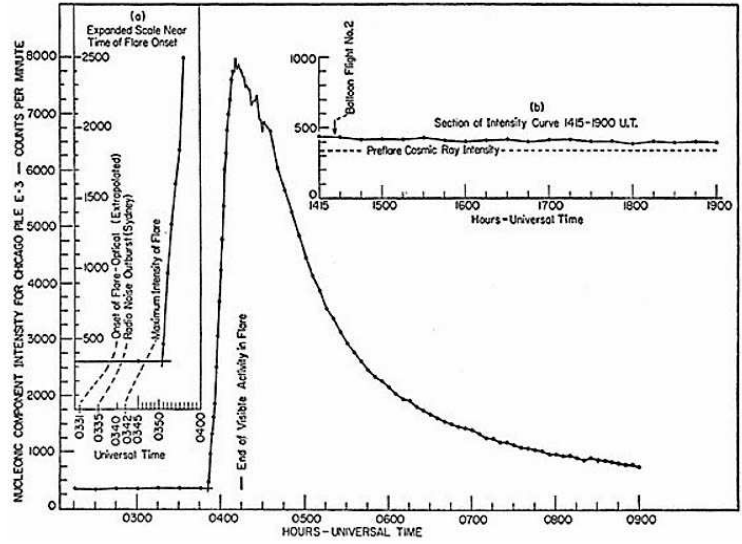}
\caption{Neutron monitor observations during 1956 solar flare event, copyright by APS \cite{PhysRev.104.768}.} \label{Fig0c}
\end{center}
\end{figure}

\begin{figure}[!t]
\begin{center}
\includegraphics[scale=0.306]{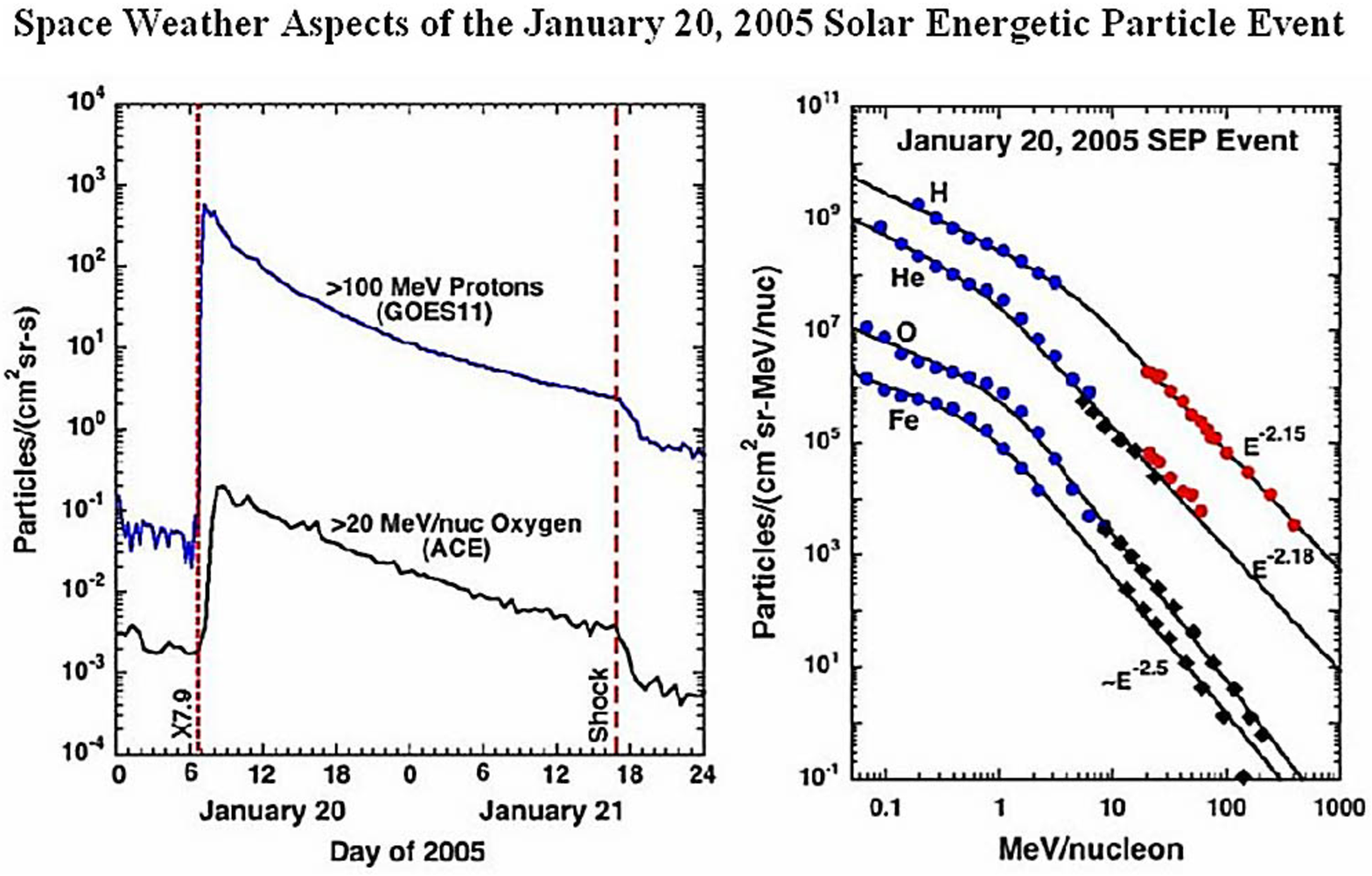}
\caption{Proton spectra during the solar flare 2005.} \label{Fig0d}
\end{center}
\end{figure}

 We shall consider a quite different, prompt, solar neutrino flare signal originated at the same solar surface where the burst occur, by proton-proton or proton-helium (or nuclei-nuclei) scattering as soon as (seconds, minutes) the flare is born.
These scattering and showering are guaranteed by the same hard and direct component of the solar flare: its neutron solar flare (see Fig.~\ref{Fig0c}).
 If their arrival and interactions on Earth atmosphere have equivalent interactions on solar corona, their expected flux has a detectable (SK) upper bound spectra. Moreover, the hard $\gamma$ spectra of solar flare, see Fig.~\ref{Fig0d} observed on Earth sky provides solid argument for a corresponding minimal hard spectra of $\nu$ secondaries in the solar corona plasma (a lower bound of $\nu$  signals). Indeed, as we shall argue in more details, the location or geometry of the particle acceleration and the consequent target solar corona density where the hitting takes place, may differ from our terrestrial one (the upper bound spectra), leading to a partial (about 5\%) suppressed output (lower bound spectra) respect to the ideal efficient scattering (ideal like in our atmosphere). Therefore, the pion trace in a hard gamma (tens or hundreds MeV) solar flare trace may be below the expected ones, based on efficient atmospheric terrestrial proton scattering. See Fig.~\ref{Fig3}. Anyway let us underline once again, the observed hard gamma component (if associated to neutral pion, of hadronic nature) is assuring  a corresponding tens or hundreds MeV neutrinos associated to the charged $\pi$;  the birth place should be at the same  HXR (hard x-ray)  foot-point where the solar flare particles jets are beamed downward to the solar corona. Near future megaton detectors for neutrino as the HK in Japan and-or in Korea (as well as IceCube upgraded Megatons detectors like PINGU) might lead to a realistic  discover of their existence\cite{2017arXiv170701987F}.
\begin{figure}[!th]
\begin{center}
\includegraphics[scale=0.306]{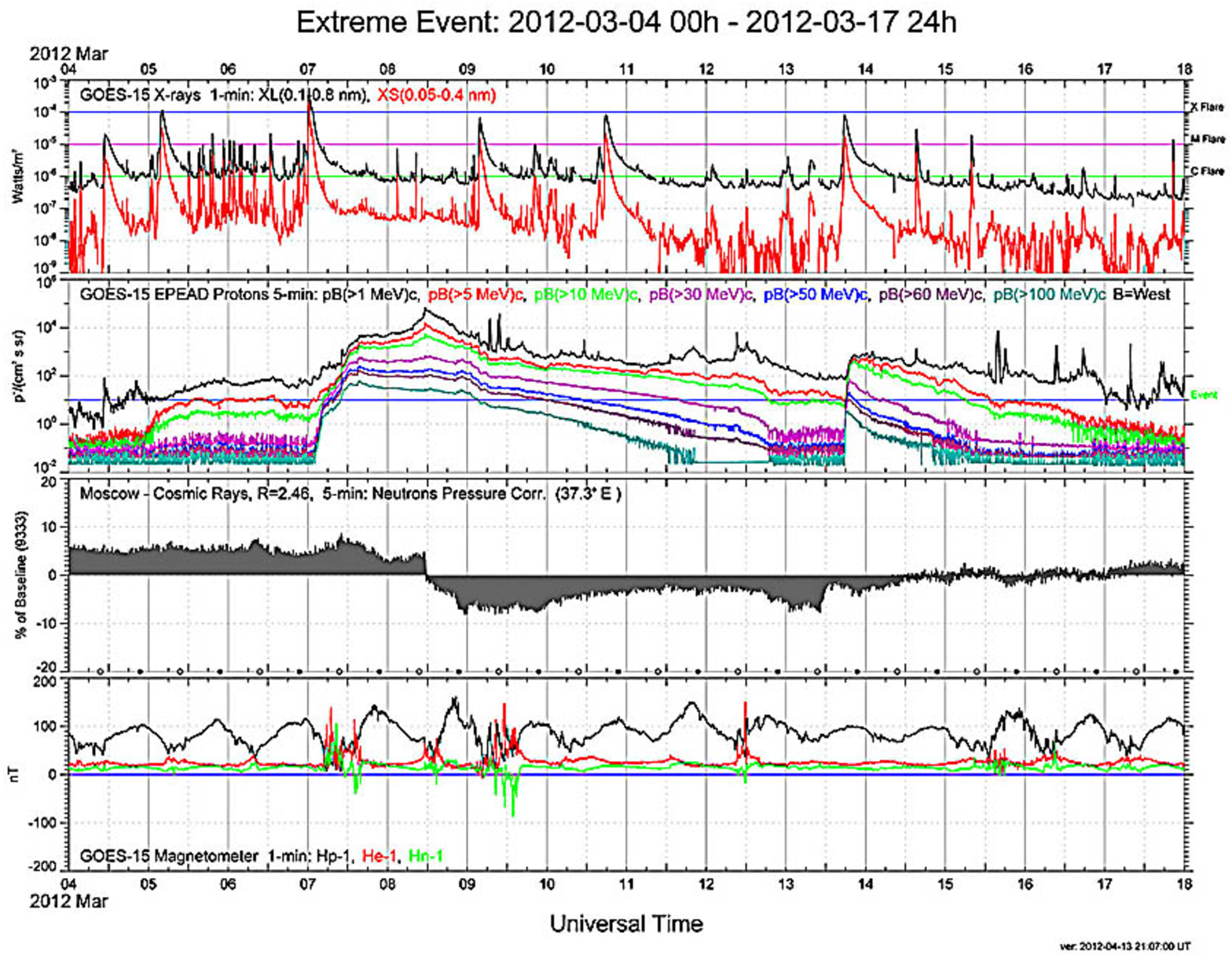}
\caption{The whole diagram of the Solar Flare 2012 with correlated x-ray flare,
Corona Mass Ejection (CME) at different energies, the Forbush late decrease observed via neutron flux, the geomagnetic storm.} \label{Fig0e}
\end{center}
\end{figure}

\section{Solar Neutrino Flare early estimate}
\begin{figure}[!t]
\begin{center}
\includegraphics[width=0.9\textwidth]{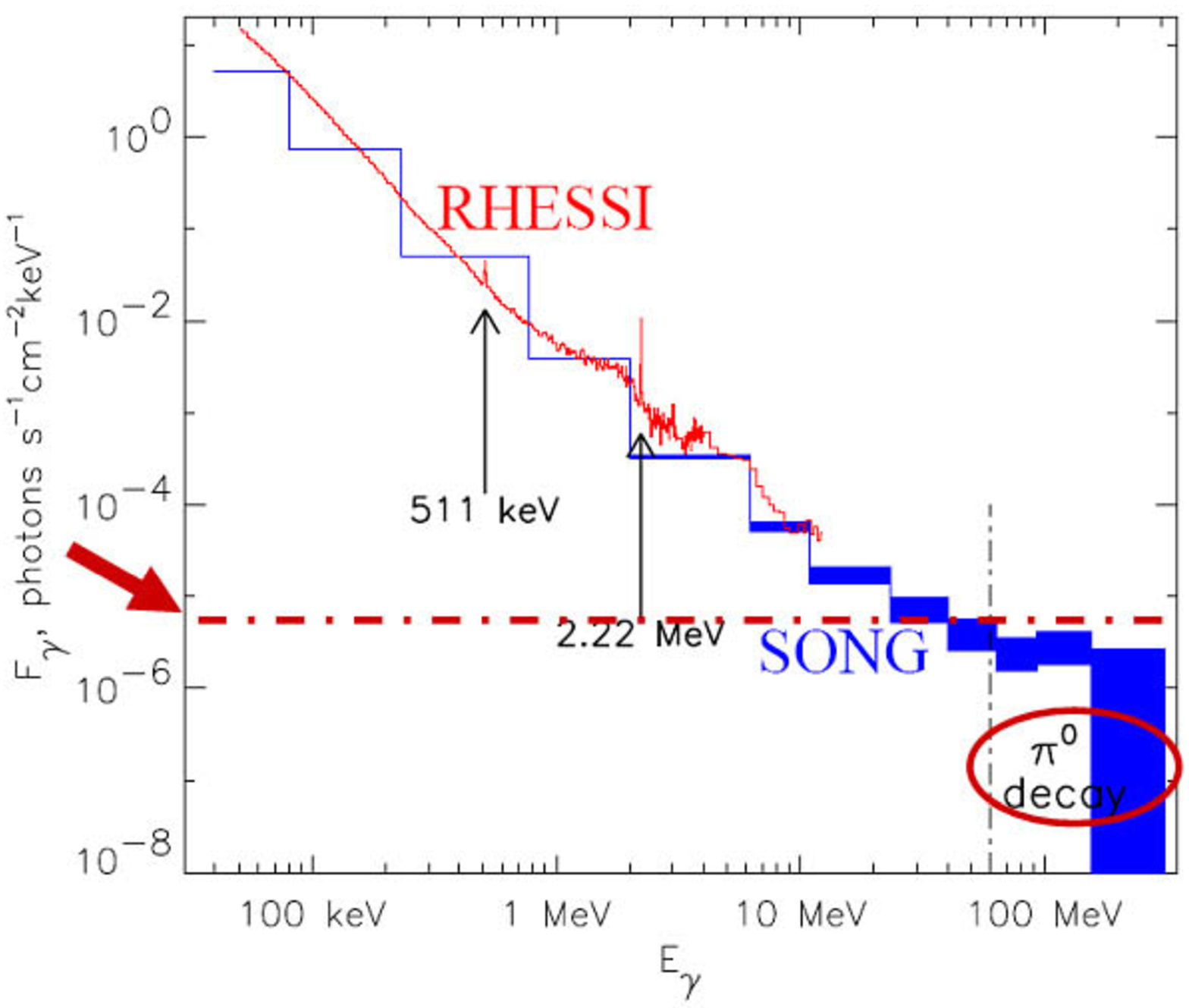}
\caption{X and gamma spectra of one of the highest energetic flare. The tens-hundred MeV photons may be originated by a nucleon or nuclei component  related to GeV hadron onto solar corona plasma  scattering range. The hard $\gamma$ component exhibit a peak flux nearly ten times the atmospheric neutrino one for several minutes. The corresponding event number it is $0.05-0.2$ event for SK ($20.000$ water tons), above unity  ($1-4$) for future HK Megaton detectors.} \label{Fig3}
\end{center}
\end{figure}
As we mentioned during the largest solar flare, of a few minutes duration, the particle flux escaping the corona eruption and hitting later on the Earth, is of $3-4$ order of magnitude above the common atmospheric CR (cosmic Ray)  background. If the flare particle interactions on the Sun corona is taking place as efficiently as they do in our terrestrial atmosphere, than their earliest proton-proton scattering, their pion  secondaries  and their muons decays ejected in solar corona are leading to a prompt neutrino fluency on Earth comparable to one day integral terrestrial atmospheric neutrino activity (upper Bound). This cumulate event number in SK is $6-8$ times the atmospheric $\nu$ per day.
One may therefore expects a prompt increase of the neutrino signals of the order of  several events made by such rarest solar flare neutrinos. Therefore, there may exist a prompt solar flare neutrino astronomy \cite{1009-9271-3-S1-75, 1126-6708-2004-06-045, 1402-4896-2006-T127-008}.
In present neutrino detectors as SK the signal is just on the edge, but, to the extent of the authors' knowledge, it has been never revealed in Super Kamiokande
\cite{1126-6708-2004-06-045, 1402-4896-2006-T127-008}. Sun density at the flare corona might be diluted and pion production may be consequently suppressed (more than one order of magnitude) leading to a signal at a few percent the expected one under the above considerations. This may be the reason for the SK null detection. Indeed the low Gamma signals recently reported \cite{Grechnev2008} confirms this suppressed signal, but just at the SK detection edge. Unfortunately the neutrino
signal at hundred MeV energies is rare while the one at ten MeV or below is polluted by Solar Hep neutrinos. Because the neutrino cross section energy thresholds (several MeV for electron, several hundreds for muons) the detection curve increases (in first approximation) with the energy square.
 Therefore, the expected
signal is not dominated at the solar Flare spectra expected maxima (30-50~MeV), but at higher ones (hundreds MeV energies). The solar Flare  $\nu$ discover that might be greatly improved by the peculiar anti-neutrino component via Gadolinium target mass, in next SK detectors \cite{1402-4896-2006-T127-008}, see Fig.~\ref{Fig6}. Our earliest \cite{1402-4896-2006-T127-008} upper limit estimate for October - November 2003 solar flares and the recent January 20th 2005 exceptional flare is leading to signal just below unity for Super-Kamiokande and to several events well above unity for any future Megaton detector.

\begin{figure}[!t]
\begin{center}
\includegraphics[width=0.9\textwidth]{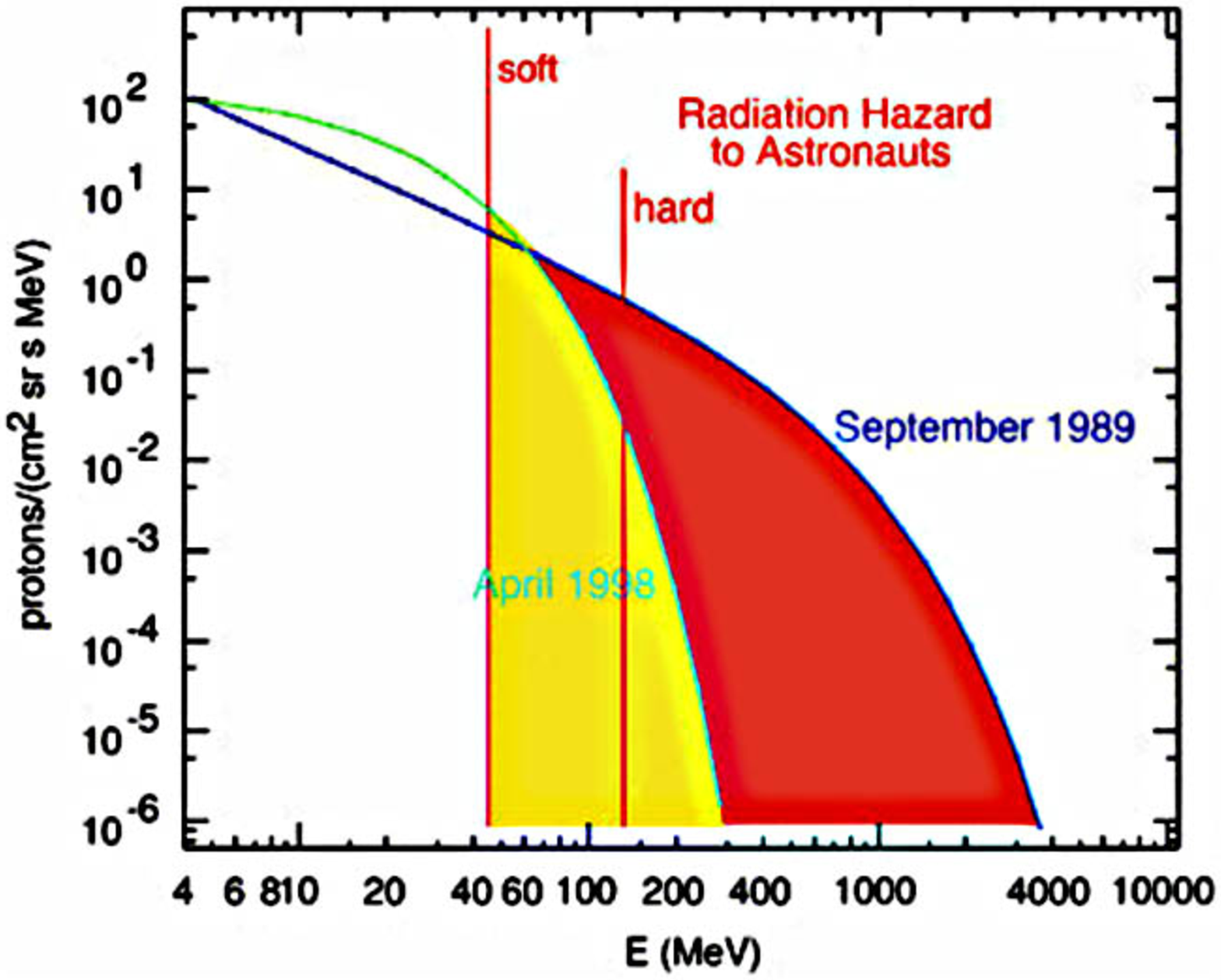}
\caption{The soft tens MeV (yellow-orange) or the harder hundreds MeV (red area) nucleon or nuclei component  of the solar flare (related to GeV hadron onto solar corona plasma  scattering range) may put in danger the astronauts long journey in space.} \label{Fig4}
\end{center}
\end{figure}

\begin{figure}[!t]
\begin{center}
\includegraphics[width=0.9\textwidth]{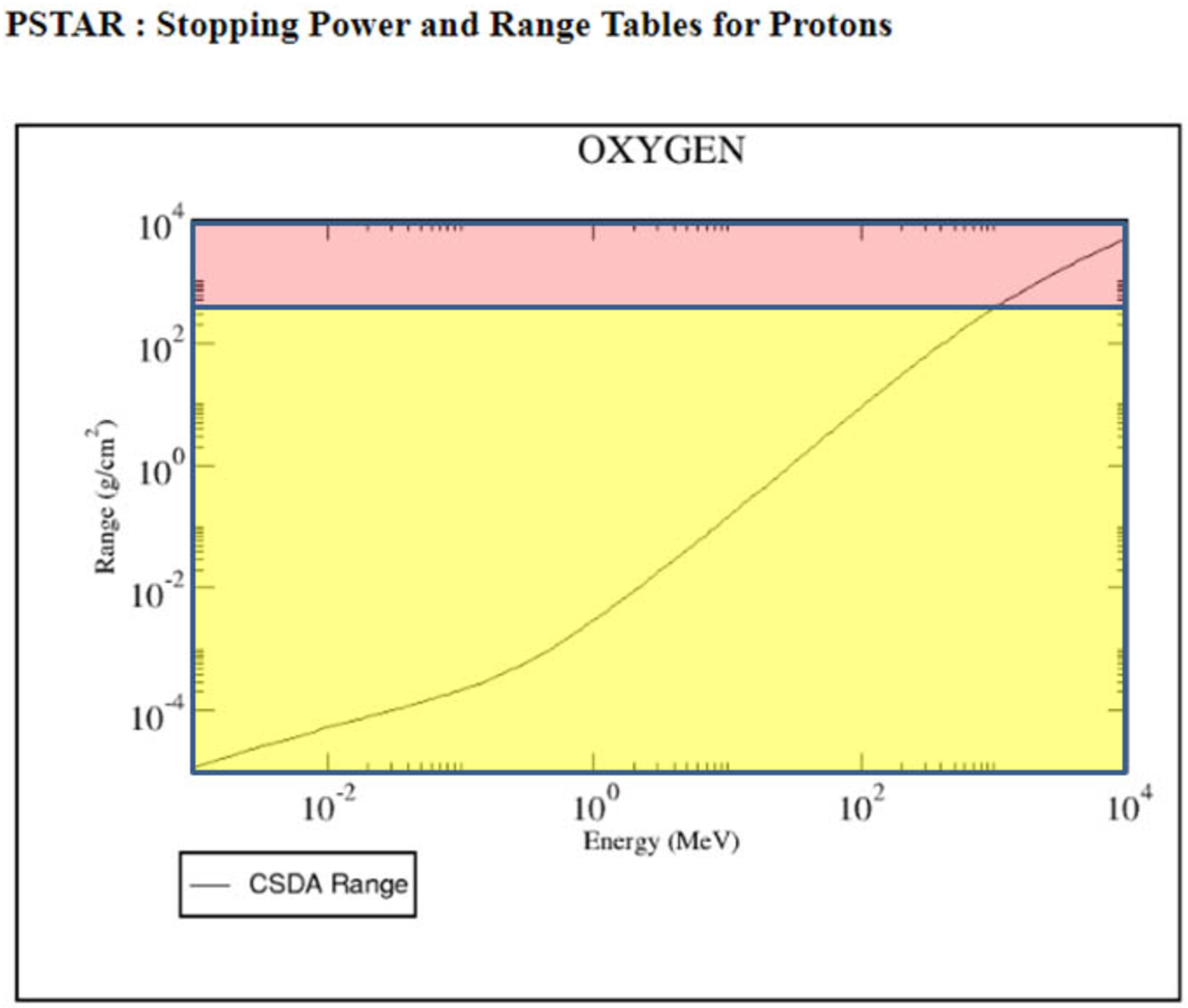}
\caption{From PSTAR web site; as above (soft tens-hundreds MeV up to hardest GeVs, in orange background) it shows the range for proton, by the energy losses and stopping power: assuming  half a meter size wall or $500$ g. $cm^{-2}$ slant depth of typical jet propeller, (Kerosene-Oxygen, or just Oxygen) it is possible to derive the corresponding energy screen edge, around 1 GeV: the yellow-orange and the red areas respectively tag the most dangerous soft (screened) and the hard (unscreened) component in highest Solar Flare events. A nominal half a meter depth guarantees a reasonable protection of most of the lethal solar flare radiation. A prompt neutrino alarm may offer the precursor precious time for astronauts needed to hide themselves in the save room} \label{Fig4}
\end{center}
\end{figure}

\section{Why it is so relevant for us: Human space journey}

One of the basic reason to have a neutrino solar flare astronomy is to use  in an independent way the
solar neutrino flavor oscillation and mixing, even taking into account the MSW effect in peculiar configurations.
However, the social argument in favor of a solar neutrino detection is more pressing: the possibility to have a prompt alarm system both for any robotic, in orbit satellite (for communication, military, scientific detection), but also and in particular for the human astronaut safety in  far space flight. Our life (better to say, most animal life) is well protected on Earth by two main shields: air atmosphere (comparable to ten meter of water wall) and terrestrial magnetic fields.  The magnetic field partially screens the most lethal soft spectra even for a huge solar flare reaching near orbit International Space Station, but astronaut in a far space flight, or in mission around our Moon are at life risk without no atmosphere or magnetic defence.
 The several years long journey to Mars makes very urgent the problem of the solar flares and their prompt detection and alert.
 Naturally there is a long list of elementary particle physics interested in the solar flare detection: their flavor mixing, the MSW oscillation inside the sun, the independent solar neutrino astronomy and the correlated plasma and nuclear diagnostic of the such huge explosions. These subject will be discussed elsewhere.

\subsection{Astronauts life defence}
 There is of course the possibility to observe a solar flare by its direct X ray flare, via satellites like SOHO or the most recent Parker Solar Probe, sent on 12th of August 2018.
 However, from these satellites as from the Earth itself  one cannot observe the whole  Sun in both sides at the same time. Therefore, hidden Solar Flare are occurring at least for a large fraction
  of the time, hidden to the X ray view. On the contrary neutrino detection is transparent (out of the mixing conversion) all along the Sun, which is transparent at hundreds-GeV energies. The solar flare nearly isotropic distribution
  is more visible than any beamed X-gamma flare jets.

 Therefore, prompt solar flare neutrinos are the fastest and the earliest and somehow best courier of the lethal out-coming particle rain flare. Moreover, their inner depth information may well reveal  interior engines in these explosions.
 The possibility to have a very rapid  alarm (by HK neutrino events) may allow the astronauts, for example, to immerse themselves immediately inside an inner large water reserve container, able to screen them from the main soft (50-100 MeV)  energy solar wind that would strike nearly half an hour later. The future Martian  mission may deeply depend on the ability to foresee, inform and protect the astronauts flight by a fast and prompt solar neutrino detection \cite{2011ICRC....6..309F}.
  One or few tons of water reserve might be too small anyway for a save cylinder swimming room.
  One of the best shields screen to solar flare radiation might be an  equipped and a  save wider  room  (maybe an inflated one)  surrounded by most of the rocket fuel (several tens of tons of chemical needed for the back return from the Mars journey): such a safe room may be located by nearly half a meter size double cylinder container in axial geometry, twin cylinder whose radius would  be nearly of 2 meters up 2.5 meters, while its height may be about 4 meters; this narrow room may guest for several hours of the flare winds of lethal danger, the $6-8$ crew astronauts. The fuel mass volume around the cylinder (with a top and a bottom coverage) is the simplest and best protection,  nearly of $30-40$ tons of the jet fuel (oxygen like chemical liquid) as the needed one for a Mars journey entire mission. The save room must be reached at the earliest time connected, as we suggest, to an ideal prompt  terrestrial neutrino alarm system, in a short time windows of $15-30$ minutes time (or twice as much for the Mars regions in arrival or departure epochs).

\section{Where and why solar neutrino flare may be soon observed}

Large size neutrino detectors (1, 2, 10 Megaton) at tens or hundreds up to GeVs  (HK, PINGU, DeepCore) are under construction; even tens GeV gamma  (and $\nu$) signals  might occur in hardest solar flare possibly detectable in ICECUBE.  The recent peculiar solar flares as the October-November 2003 and January 2005  \cite{1402-4896-2006-T127-008}  were source of high energetic charged particles: large fraction of these primary particles, became a source of both neutrons \cite{1126-6708-2004-06-045} and secondary kaons $K$, pions $\pi^{\pm}$ by their particle-particle spallation on the Sun surface. Consequently,  final secondaries muon and electronic neutrinos and anti-neutrinos, should be released by the chain reactions occurring on the sun atmosphere. The solar flare neutrinos reach the Earth with a well defined directionality and within a narrow time range. Their corresponding average energies  suffer negligible energy losses. The opposite occur to downward flare. In the simplest approach, the main source of pion production is the common scattering
$p+p$ making a $\Delta^+$ mostly at its center of mass (of the resonance) whose mass value is
$m_{\Delta^+}=1232$~MeV. As a first approximation and as a useful simplification after the needed boost of the secondaries
energies one may assume that the total pion Delta energy is equally distributed, in average, in all its four final remnants each at an average 30~MeV energy. An additional relativistic boost may lead to higher energies, partially beamed. The consequent spectra for largest solar flare  are described with the corresponding future detector mass (1-2 Megaton) threshold  respectively for HK in Japan, HK in Korea, before their mixing in arrival on Earth; PINGU and DeepCore in IceCube,  might reach 5 or ten megatons as well a twice or five times larger rate than the foreseen one for the two Megaton case. Their mixed signal are described in Fig.~\ref{Fig5}. Their mixed and filtered signal are also described in Fig.~\ref{Fig6}. In an anti-neutrino electron mode obtained by a filtering detector (like the Gadolinium one) the absence of the solar neutrino noise it is totally excluded and the solar flare signal is in a sharpest signal-noise mode.

\begin{figure}[!t]
\begin{center}
\includegraphics[width=0.9\textwidth]{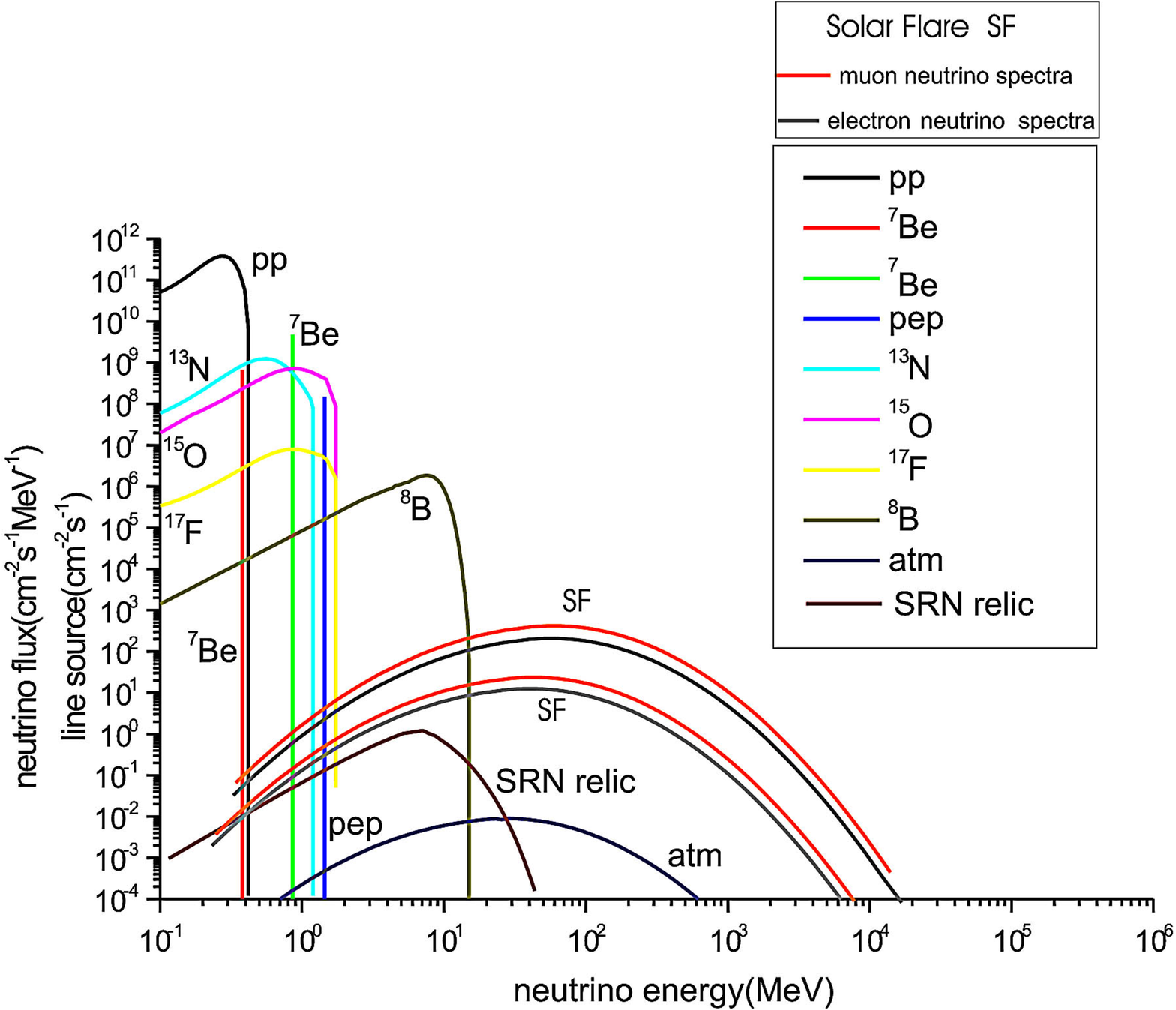}
\caption{The whole  neutrino spectra with the main solar neutrino noise. The different flavor energy threshold is no  marked: it appears respectively for electron, muon, tau flavors around energy 5~MeV, 400~MeV, 10~GeV. The hardest ones are advantaged to be detect by larger cross section; the eventual GeV muon neutrino signal is amplified  by its track: external trough going muon may enlarge the detection volume of the solar flare even doubling the observing volume.} \label{Fig1c}
\end{center}
\end{figure}
\begin{figure}[!t]
\begin{center}
\includegraphics[width=0.9\textwidth]{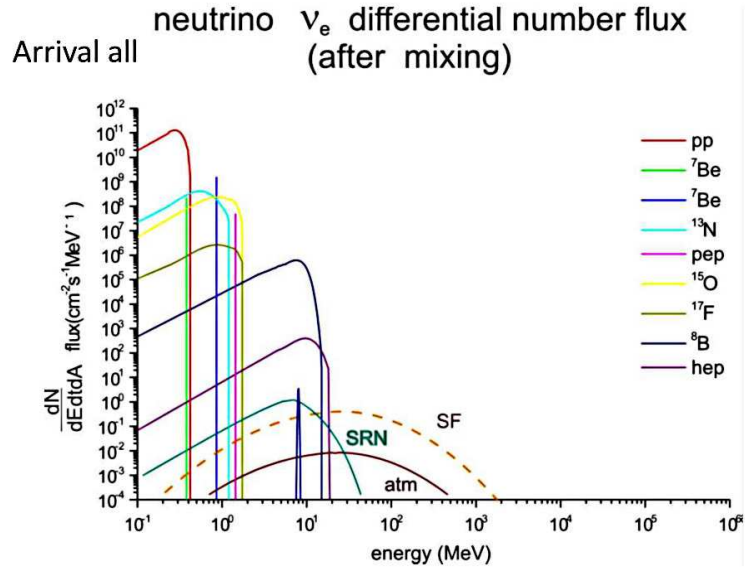}
\caption{The whole muon anti-neutrino electron (comparable to the anti neutrino muon and tau ones) spectra with the main solar neutrino component almost absent. The upward and downward  atmospheric muon neutrinos would show the suppression of factor half  due to average mixing into the tau flavor\cite{Farg2004NPB, Farg2007NPB}. The absence of solar neutrino noise makes antineutrino detection an  ideal tool to disentangle the solar flare signal. Because the thresholds in the neutrino cross-sections the signals may rise at hundred MeVs.} \label{Fig5}
\end{center}
\end{figure}

\begin{figure}[!t]
\begin{center}
\includegraphics[width=0.9\textwidth]{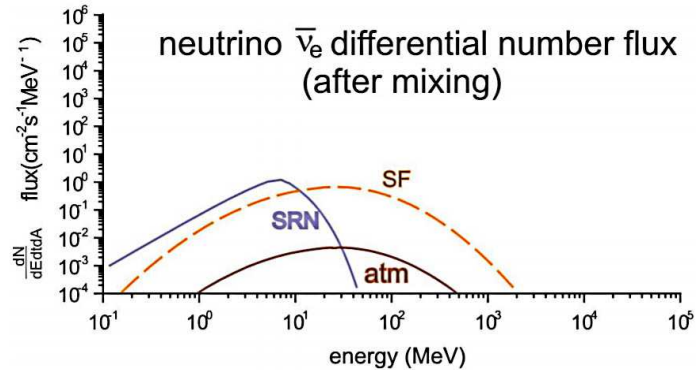}
\caption{The anti-neutrino electron (comparable to the anti neutrino muon and tau ones) spectra with the main solar neutrino component almost absent. The absence of solar neutrino noise makes antineutrino detection an  ideal tool to disentangle the solar flare signal. Because of energy  thresholds in the birth and in neutrino cross-sections the signals may rise at tens or better hundred MeVs} \label{Fig6}
\end{center}
\end{figure}

\begin{figure}[!t]
\begin{center}
\includegraphics[width=0.9\textwidth]{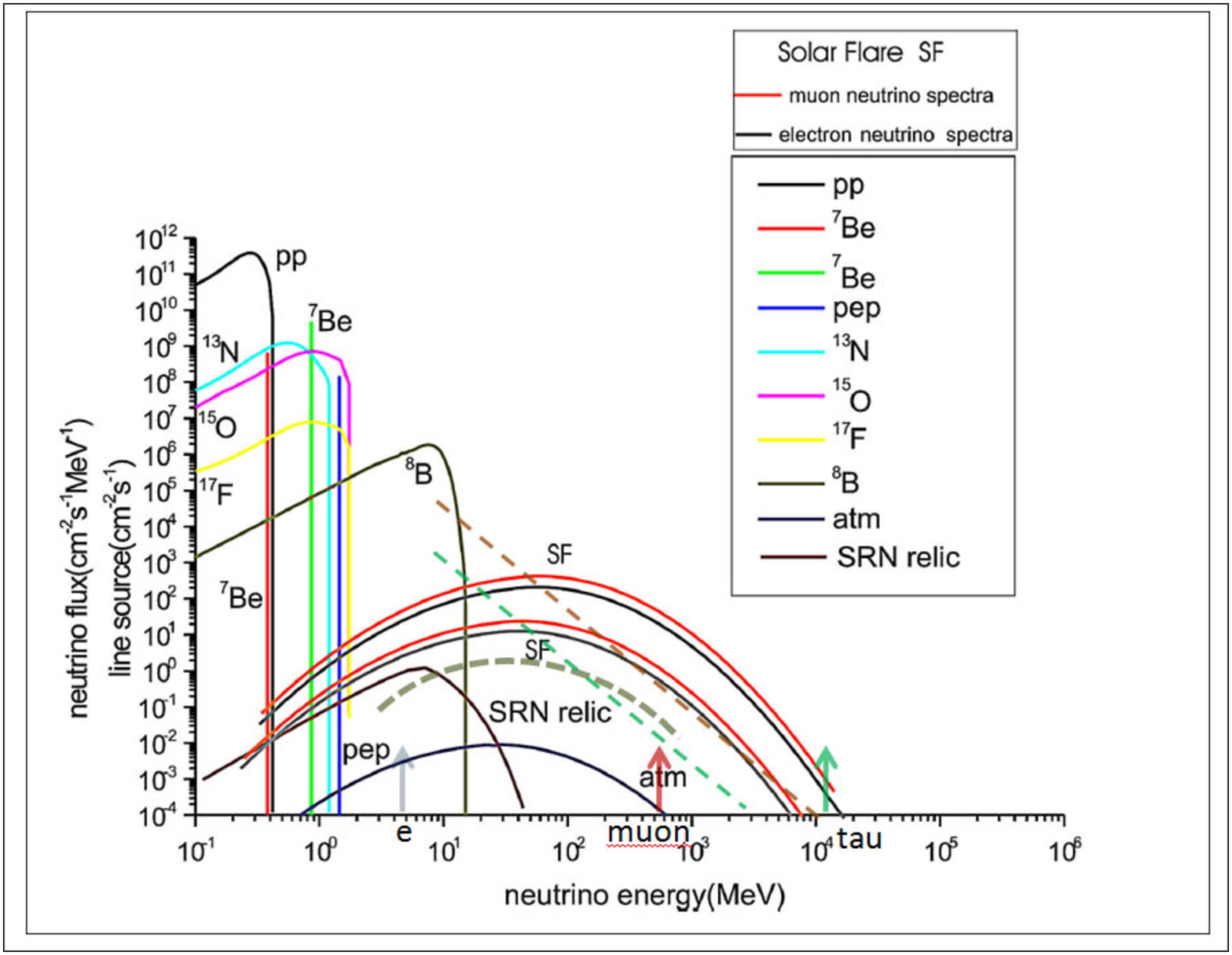}
\caption{The neutrino electron, muon and tau  spectra. The detection energy  thresholds for each flavor it is marked by an arrow (few Mev, hundreds MeV, 10 GeV) respectively. The neutrino cross-sections growth with the energy lead to a detection line respectively for SK and HK at the edges of solar flare discover. Its best appearance is above 30 MeV, better possibly at hundreds MeVs. The dashed red line describe the SK expected ability to detect the brightest Solar Flare.
However the more realistic, less luminous (gray dashed parabola) Solar Flare might be observed in future Megaton detectors as HK by the corresponding more wider mass volumes (green dashed line). } \label{Fig6}
\end{center}
\end{figure}

\subsection{Solar neutrino flares above GeV energy: the muon track amplifier}

The near future and the present Megaton neutrino detector might be able to reveal the
solar neutrino flare, as shown in Fig.\ref{Fig0}. One may note that the imprint of the anti neutrino electron
 is much clean because of the silent solar thermonuclear emission or noise in Fig.\ref{Fig6}. This imply that also Gadolinium enhanced Megaton detector may play a role in solar flare and possibly in Relic SN neutrinos signals.  Above the GeV energy the appearance of the first solar flare muon neutrinos will be a revolutionary one: (out of the very recent ICECUBE connection on $22$th September 2017 of a hard TeV event track with a probable gamma AGN source,the celebrated TXS 0506+056)  it has never been observed any muon neutrino, by any star, yet. Highest energy (PeVs-EeV) upward tau neutrino astronomy escaping the Earth by their in flight tau air-shower, it is just beyond the corner; \cite{Fargion(2002)}, \cite{2011ICRC....6..309F},\cite{Farg2007NPB} .   Moreover, there is a remarkable enhanced effect for GeV muons.  The muon track in water extends nearly 5 meters for each GeV of energy, almost in linear growth. The possibility to observe along the edges of the SK or HK container external muons (through going muons) that are pointing to the sun during the flare may increase the effective volume by a significant factor. For instance at 10 GeV the eventual solar flare may double the SK volume and (because of the 2.5 rock density), it may even make three times a larger detection volume. The same amplify effect in HK is less remarkable because of the wider HK container sizes; nevertheless, at several GeV muon tracks may amplify solar neutrino flare detection by a large fraction both in SK and HK detector by external masses.

 \section{Conclusions}
The solar flare neutrino is at hand. Its detection is well within  HK, Deep Core, PINGU detector systems.
The absence of detection in SK may be due to a low mass density of flare scattering in the outer flare propagation mass ejection and the reduced component of jet hitting back the coronal solar surface.
However, there is also a prompt gamma signal that imply a neutrino one that must be sooner or later observed in largest Megaton neutrino detectors. Its discover will offer the first nearest verified neutrino hadronic imprint in modern neutrino Astronomy.

\section{Acknowledgement}
The work by M. Khlopov was supported by grant of the Russian Science Foundation (project No-18-12-00213).
\section{Dedication and remembrance}
The present article is devoted to the memory  of the   promulgation of the Italian racial law occurring
on  the $5$th September 1938, today at $80$th  anniversary. Even with greatest gratitude for the all of the hundreds of heroic Italian rescuer of Jews (like Giorgio Perlasca), nearly a third of the Italian Jewish population has been first excluded and later deported to Germany, almost all with no return. It was like that a third of the Italian population, today twenty million, was stolen of dignity and later, of life.

\bibliographystyle{ws-procs9x6} 

\end{document}